\documentclass[a4paper,11pt]{article}

\usepackage{pos}
\usepackage{lipsum} 
\usepackage{color}

\usepackage{setspace}
\usepackage{titlesec}
\titlespacing{\section}{0pt}{*1.0}{*0.66} 
\titlespacing{\subsection}{0pt}{*1.0}{*0.67} 
\title{First Ultra High Energy Neutrino Search with a Hybrid Phased and Traditional Detector in the Askaryan Radio Array}

\ShortTitle{First UHE Neutrino Search with a Hybrid Phased and Traditional Detector in ARA}

\author*[a]{Paramita Dasgupta} 
\onbehalf{for the ARA Collaboration \\{\normalsize \normalfont(a complete list of authors can be found at the end of the proceedings)}\\}

\affiliation[a]{Dept. of Physics, Center for Cosmology and AstroParticle Physics,\\
The Ohio State University, Columbus, OH 43210}

\emailAdd{dasgupta.80@osu.edu}

\abstract{
\sloppypar
\hyphenpenalty=10000
\exhyphenpenalty=10000
\tolerance=1000
\emergencystretch=1em
The Askaryan Radio Array (ARA) is an in-ice ultrahigh energy (UHE, >10 PeV) neutrino experiment at the South Pole, designed to detect neutrino-induced radio emission in ice. It consists of five independent stations, each featuring a cubic lattice of in-ice antenna clusters spaced ~30 m apart and buried ~200 m below the surface. The fifth ARA station (A5) is unique due to its central phased array string, which employs an interferometric trigger to enhance sensitivity to weak signals otherwise buried in noise. This low-threshold trigger makes ARA the first in-ice radio neutrino experiment to demonstrate a significant improvement in detecting low signal-to-noise ratio (SNR) radio signals.\\
We present progress toward the first UHE neutrino search utilizing A5's hybrid detection capability, incorporating advancements in data selection and background rejection. This analysis is the first to fully apply dedicated event selection to both components of ARA's hybrid detector, improving directional reconstruction and significantly enhancing background rejection compared to previous analyses. This approach paves the way for next-generation in-ice UHE neutrino experiments.

}

\ConferenceLogo{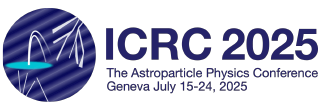}

\FullConference{39th International Cosmic Ray Conference (ICRC2025)\\
 15–24 July 2025\\
Geneva, Switzerland\\}

\begin{document}

\maketitle

\section{Introduction}\label{sec1}

\par 
Over the past few decades, ultra high energy (UHE, $>$10~PeV) astrophysics has become a critical frontier for exploring the most extreme and distant cosmic environments. UHE cosmic rays (UHECRs) with energies above 1~EeV interact with cosmic microwave background (CMB) photons, resulting in Greisen-Zatsepin-Kuzmin (GZK) suppression around 100~EeV~\cite{Greisen:1966jv,Zatsepin:1966jv}. This limits their propagation to $\sim$50~Mpc, reducing the observable flux at Earth to roughly one particle per square kilometer per century, while magnetic field deflections obscure directional information.
\par
In contrast, neutrinos, being electrically neutral and weakly interacting, can travel cosmological distances undisturbed, preserving directional and spectral information from their sources. Because they can penetrate deeply into otherwise opaque astrophysical environments, neutrinos provide unique insights into the inner acceleration regions and structural properties of cosmic sources, not just their edges. UHE neutrinos may be produced directly in astrophysical accelerators or via the GZK mechanism. The Askaryan effect~\cite{Askaryan:1961pfb} enables detection of these neutrinos through coherent radio Cherenkov emission in dense media such as Antarctic ice. 
\par
The Askaryan Radio Array (ARA) employs this technique by deploying in-ice radio antennas to instrument large volumes with kilometer-scale radio attenuation lengths, allowing cost-effective monitoring for neutrino interactions. Similar techniques were employed by the ANITA experiment~\cite{Prohira:2018mmv}, which flew balloon-borne radio antennas over Antarctica to detect radio signals from neutrino interactions in the Antarctic ice sheet. ARA deployed its first station in 2012 and has since enhanced its sensitivity with a beamforming trigger system in its fifth station, lowering detection thresholds and improving sensitivity to weak, broadband impulsive signals.
\par
This contribution presents the first UHE neutrino search using ARA’s hybrid detector system, which combines a traditional antenna cluster with a beamforming phased array string. It is the first UHE neutrino search using this hybrid dataset, leveraging the phased array trigger to collect data from both detector components. The methods and results from this analysis highlight the potential of hybrid systems and inform the design of next-generation experiments, including IceCube-Gen2~\cite{IceCube-Gen2:2020qha}.

\section{ARA Detector Description and Experimental Data Set}\label{sec2:detector}
\par
The Askaryan Radio Array (ARA) comprises five autonomous detector stations deployed in a hexagonal grid with 2~km spacing, optimizing the effective area for neutrino energies above $10^{18}$~eV while minimizing instrumentation requirements. Each station has four ``measurement strings'' installed in vertical boreholes drilled to a depth of approximately 200~meters. Each string houses two vertically-polarized (VPol) and two horizontally-polarized (HPol) antennas, all sensitive to radio frequency (RF) signals in the $150$–$850$~MHz band~\cite{ARA:2019wcf}.

\begin{figure}[htbp!]
  \centering
  \includegraphics[width=0.45\textwidth]{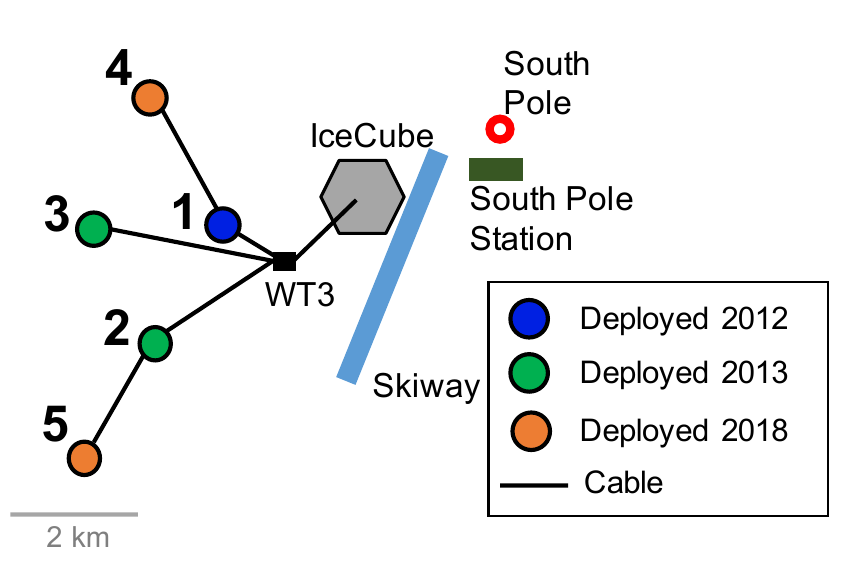}
  \hfill
  \includegraphics[width=0.45\textwidth]{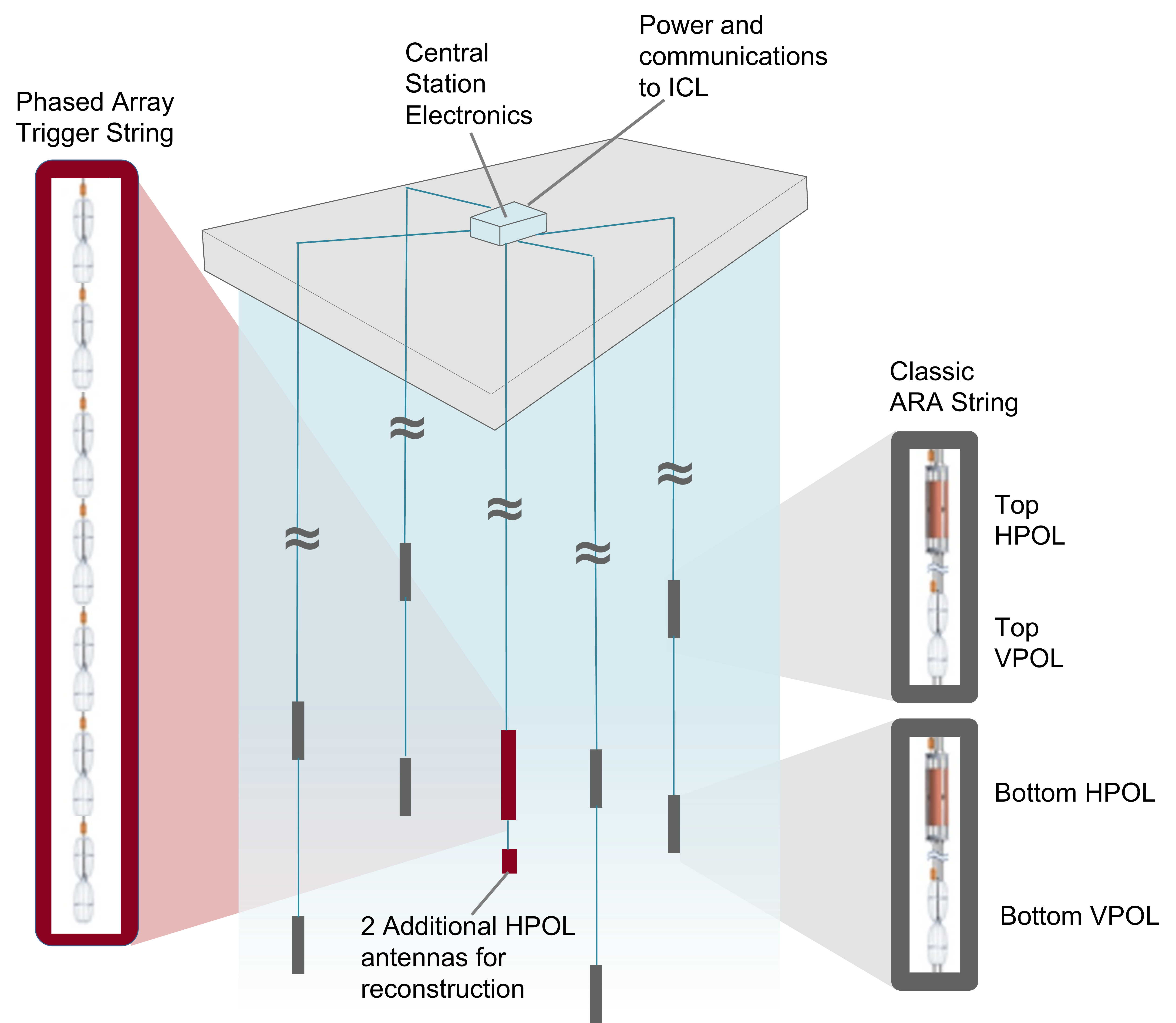}
  \caption{Left: Layout of the ARA station array and deployment timeline. Right: Layout of the two subdetectors at ARA’s fifth station (A5). Traditional ARA antennas are shown in grey; phased array antennas, closely spaced along a central string, are shown in red. All other stations (A1–A4) share the same traditional antenna layout but do not include a phased array string.}
  \label{fig:A5_detector}
\end{figure}
In addition to measurement strings, each station includes two calibration strings deployed at similar depths, typically $\sim$40 meters from the station center. Each calibration string contains one HPol and one VPol transmitter antenna capable of emitting broadband RF pulses or noise for in-situ geometry and timing calibration. Stations operate at an average trigger rate of approximately 6 Hz, including 1 Hz from calibration pulsers and 1 Hz of forced triggers to monitor ambient noise and detector performance. All five ARA stations (A1–A5) (Fig.~\ref{fig:A5_detector}, left panel) are fully calibrated through dedicated digitizer and in-ice antenna position calibrations ensuring accurate reconstruction of signal direction and timing~\cite{ARA:2021wmr}.
\par
The fifth ARA station (A5) is unique, featuring two integrated subdetectors: the traditional ARA array described above and an additional ``phased array'' (PA) detector (Fig.~\ref{fig:A5_detector}, right panel). The PA consists of a single string of nine closely spaced antennas (seven VPol and two HPol), deployed at a depth of approximately 180~m. This hybrid setup, combining traditional and phased array antennas, has directly influenced the design of next-generation UHE neutrino detectors, such as the proposed radio component of IceCube-Gen2.

\subsection{The Hybrid Detector System of ARA Station 5}\label{sec3:pa_trigger}
\par
The A5 station is unique as it integrates two subdetectors, each with its own data acquisition (DAQ) and trigger system~\cite{Allison:2018ynt}. The phased array (PA) subdetector uses an interferometric trigger that coherently sums signals from its seven VPol channels using predefined timing delays corresponding to 15 plane wave arrival directions, referred to as ``beams.'' Impulsive signals aligned with these directions add coherently, while thermal noise remains incoherent across beams. The PA issues a trigger when the power in any beam exceeds a threshold within a 10~ns window, maintaining a global trigger rate of approximately 11~Hz. This beamforming trigger significantly improves sensitivity to low SNR signals, enhancing the detector’s ability to observe faint neutrino-induced events~\cite{ARA:2022rwq}.

\par
Initially, the traditional ARA and PA subdetectors (see Fig.~\ref{fig:A5_detector}, right panel) operated independently. In 2020, six VPol channels from the traditional antenna cluster were integrated into available inputs on the PA DAQ, creating a unified system. The analysis presented in this paper uses data from this configuration, referred to as the A5/PA hybrid system, collected during 2020–2021. First implemented by ARA, this hybrid design serves as a prototype for future UHE neutrino detectors that combine phased array and traditional detection techniques within a single system.

\subsection{Advantages of a Hybrid Detection Approach}\label{sec3}
\par
A previous UHE neutrino search with data recorded only by the PA subdetector demonstrated that an interferometric trigger improves both trigger and analysis efficiency~\cite{ARA:2022rwq}. Building on this, the A5/PA hybrid system combines the PA instrument with the traditional ARA antennas, enhancing sensitivity and signal reconstruction.

\begin{figure}[htbp!]
  \centering
    \includegraphics[width=0.90\textwidth]{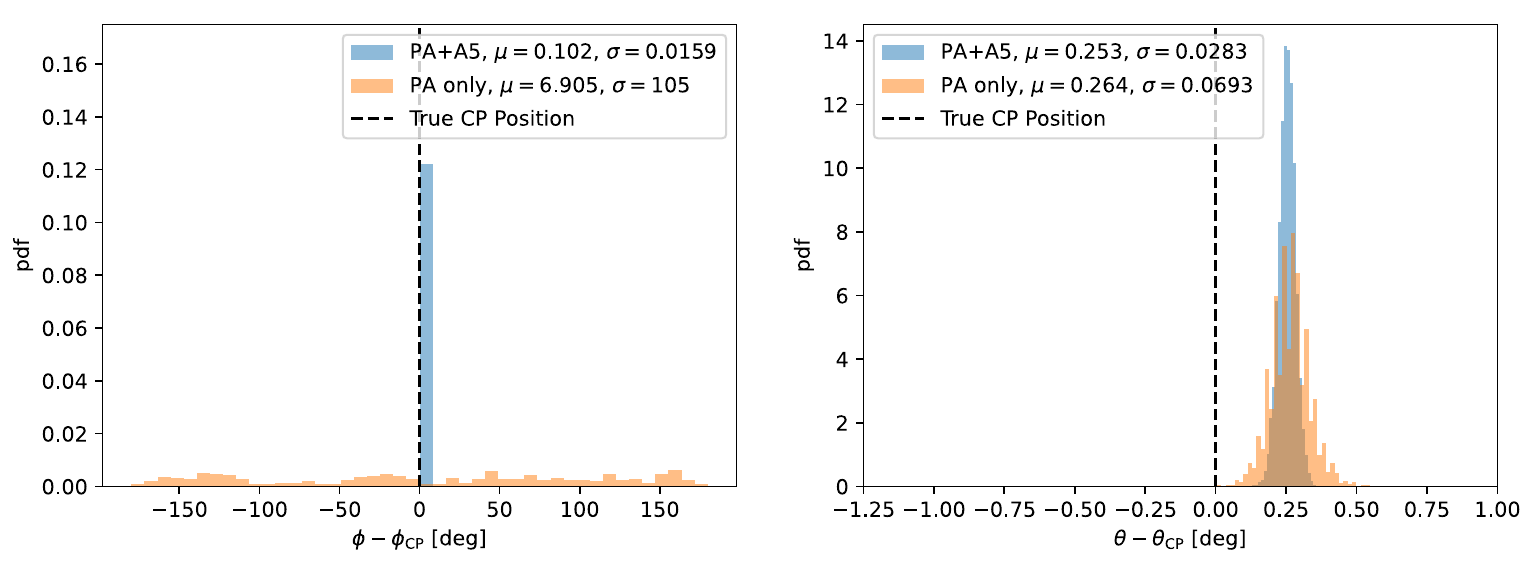}
    \caption{Reconstructed $\phi$ (left) and $\theta$ (right) of a calibration pulser relative to the true position (dashed line). The A5/PA hybrid system (blue) shows improved angular reconstruction in both $\phi$ and $\theta$ compared to the PA-only configuration (orange).}
    \label{fig:a5paReco}
\end{figure}
\par
The PA-only system has very limited azimuthal resolution due to the PA antennas being arranged on a single vertical string (see Fig.~\ref{fig:A5_detector}, right), resulting in a standard deviation of approximately 105$^\circ$ in the reconstructed $\phi$ distribution (Fig.~\ref{fig:a5paReco}, left). Adding the traditional antennas breaks this azimuthal symmetry, significantly improving the resolution to a standard deviation of about 0.016$^\circ$, as shown in Fig.~\ref{fig:a5paReco} (left). The zenith angle reconstruction is also improved by roughly a factor of two (Fig.~\ref{fig:a5paReco}, right) for high-SNR events. This enhanced angular precision greatly improves the detector’s efficiency in UHE neutrino searches by aiding the rejection of anthropogenic backgrounds clustered in azimuth. These improvements, first demonstrated by ARA, highlight the advantages of a hybrid detection approach for future UHE neutrino experiments.

\subsection{Livetime}\label{livetime}

\par
This analysis uses data collected by the A5/PA hybrid system over two years, totaling $504$ days during 2020 and 2021. The dataset starts after integrating the traditional ARA VPol channels into the PA DAQ, marking the beginning of the hybrid configuration. To ensure an unbiased analysis, we employ a blind approach~\cite{Klein:2005di}, randomly selecting 10\% of the full dataset as a representative sample. Because this subset is randomly chosen, it captures the time-dependent noise variations expected in the full dataset and is used to develop quality cuts and background rejection techniques.

\section{Simulation}\label{sec3:simu}

\par
We simulate neutrino interactions and detector response using the ARA Monte Carlo framework, \verb|AraSim|~\cite{arasim}, incorporating recent improvements for greater accuracy. One major improvement in \verb|AraSim| is the implementation of data-driven noise models constructed from forced trigger events (see Section~\ref{sec2:detector}). Additional improvements include updated signal chain gain models that incorporate amplifier properties and thermal noise based on local ice temperature, as well as refined antenna gain patterns derived from recent anechoic chamber measurements of all antenna types: horizontally polarized quadslot antennas and vertically polarized dipole antennas at the top and bottom of each string. This modified version of \verb|AraSim| simulates both the traditional antennas and the phased array trigger string~\cite{ARA:2014fyf}.

\par
Simulated events are generated according to a power-law spectrum proportional to $\phi \propto E^{-1}$~\cite{Kotera:2010yn}, with isotropic arrival directions and vertex locations uniformly distributed within 8~km of the detector. The phased array (PA) trigger efficiency is estimated based on the event’s observed SNR and the deviation of the signal emission angle from the Cherenkov angle. For each event, a random number between 0 and 1 is drawn; if this number is less than or equal to the estimated trigger efficiency at that SNR, the event is considered triggered. This trigger model reproduces in-situ performance within 15\%.

\section{First UHE Neutrino Analysis with a Hybrid Instrument}\label{sec3}
\par
This section presents the first UHE neutrino analysis using the A5/PA hybrid system of ARA, the first in-ice radio detector to combine traditional ARA channels with a phased array trigger. Building on the blind analysis framework described in Section~\ref{livetime}, we use the designated 10\% sample to develop event selection criteria that efficiently discriminate potential neutrino candidates from backgrounds. The optimized selection is then applied to the remaining 90\% of the data to set limits on the diffuse UHE neutrino flux. The methods and results from this pioneering analysis demonstrate the utility of a hybrid system and pave the way for future radio neutrino experiments. The following sections provide an overview of the analysis methodology.

\subsection{Background}\label{sec:backgrounds}

\par
\begin{figure}[htbp!]
  \centering
  
  {\includegraphics[width=0.480\textwidth]{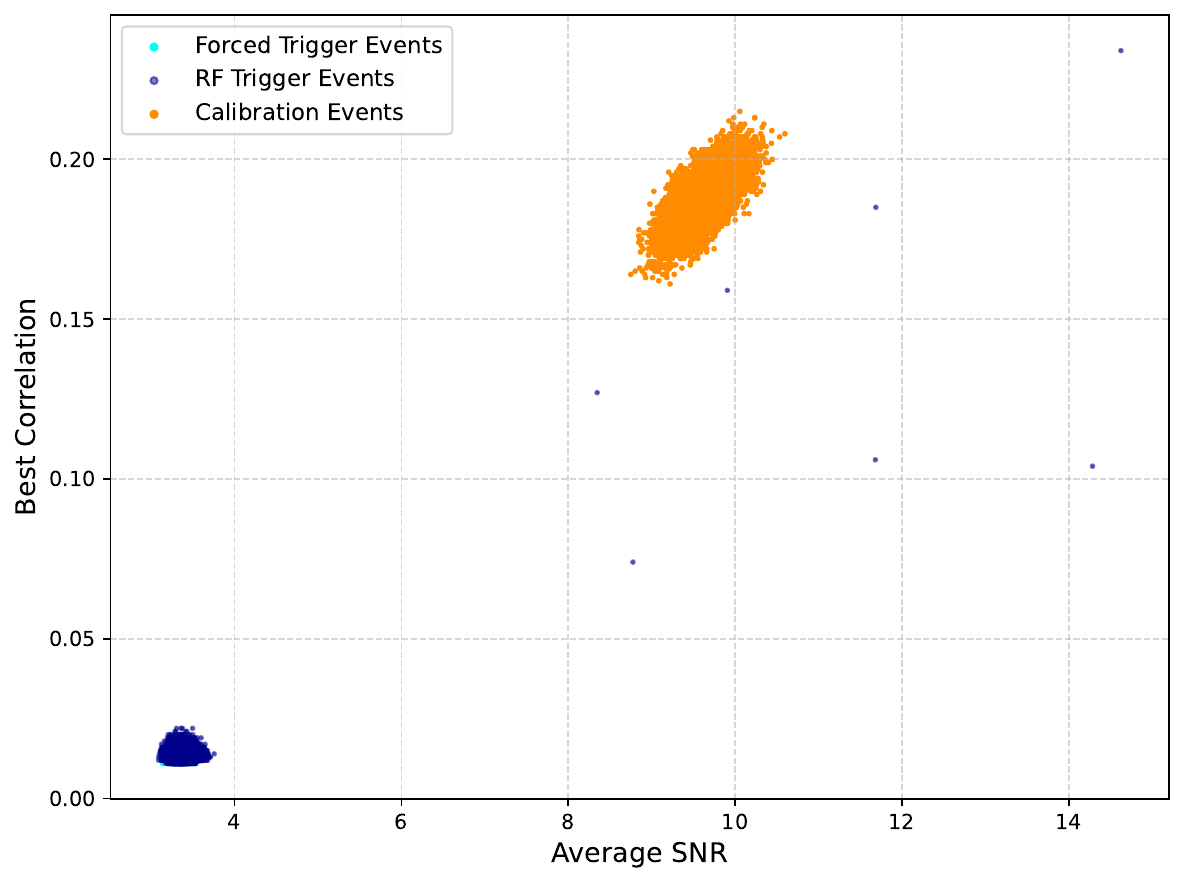}}
  \hfill
  {\includegraphics[width=0.480\textwidth]{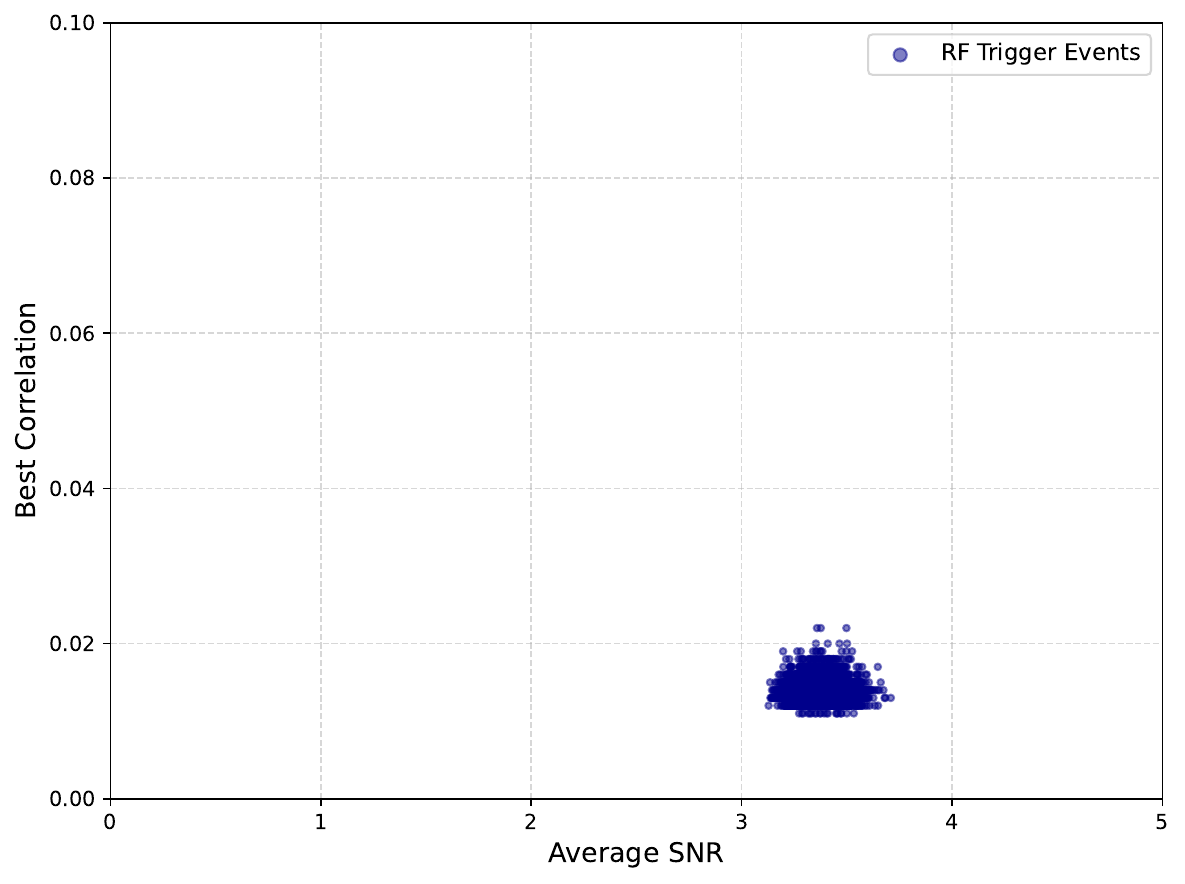}}
   \caption{Left: Distribution of all three trigger types in a subset of A5/PA hybrid data from the analysis livetime: calibration pulser events (orange), RF-triggered events (blue), and forced trigger events (cyan), overlapping with the RF-triggered distribution. Right: RF-triggered events remaining after removing calibration pulser events, forced triggers, and impulsive backgrounds, dominated by thermal noise.}
  \label{fig:data_dist}
\end{figure}

Although ARA is located at the remote South Pole, far from most human activity, the dataset remains dominated by backgrounds. To reliably identify rare UHE neutrino candidates, a thorough understanding of these backgrounds is essential before unblinding the full dataset. For characterization and mitigation, we use the 10\% blinded sample described in Section~\ref{livetime}. Tagged forced trigger events (Fig.~\ref{fig:data_dist}, left, cyan points), which capture snapshots of the noise environment independently of RF triggers (Fig.~\ref{fig:data_dist}, left, blue points), are removed as they do not originate from ambient noise fluctuations in the ice.

The remaining backgrounds fall into two broad categories: impulsive and non-impulsive, each described in the following sections along with dedicated rejection techniques designed to preserve potential signal candidates.

\subsubsection{Impulsive Backgrounds}\label{impulsive}

\par
Dominant impulsive backgrounds arise from calibration activity, anthropogenic radio signals, and cosmic ray-induced signals. Tagged calibration pulser events (Fig.~\ref{fig:data_dist}, left, orange points) are removed using timestamp information, while untagged calibration pulser events are rejected via a geometric cut of $\pm 3^\circ$ in zenith and azimuth around the known in-ice calibration pulser location (see Section~\ref{sec2:detector}). Since untagged calibration pulser events typically reconstruct within $1^\circ$ of the true pulser position, this cut efficiently removes all calibration pulser events from the 10\% sample. 
\par
Anthropogenic radio signals, typically originating above the ice surface, manifest as high average SNR outliers distinctly separated from the dominant low average SNR thermal noise distribution in the RF-triggered event dataset (Fig.~\ref{fig:data_dist}, left, blue points). These events cluster in time and space above the station and are removed using spatio-temporal clustering cuts.
\par
Cosmic ray air showers produce broadband impulsive signals via geomagnetic and Askaryan emission~\cite{DEVRIES201696} that can mimic neutrino signals. Since neutrino signals originate within the ice, all downward-going impulsive events originating above the station are assumed to be cosmic rays and are removed, resulting in a $\sim$10\% reduction in signal efficiency.

\subsubsection{Non-Impulsive Backgrounds}\label{non_impulsive}
\par
The primary non-impulsive backgrounds in this analysis are continuous wave (CW) contamination and residual RF-triggered thermal noise events (Fig.~\ref{fig:data_dist}, right, blue points) that remain after removing the impulsive RF-triggered events described in Section~\ref{impulsive}. CW backgrounds arise from narrow-band signals at known frequencies, most notably around $410$~MHz from weather balloons launched twice daily at the South Pole and approximately $210$~MHz from satellite communications. To suppress these spectral features, we apply a sine-subtraction filter originally developed by the ANITA Collaboration~\cite{ANITA:2018vwl}.

\par
Thermal noise events, triggered by random RF fluctuations from the ambient ice and detector electronics, are especially prevalent in this dataset due to the use of a beamforming trigger in the hybrid system. While the phased array trigger improves sensitivity by lowering the effective SNR threshold and increasing the likelihood of capturing weak UHE neutrino signals, it also increases the rate of low-SNR thermal events compared to other ARA stations that use higher trigger thresholds. These near-threshold thermal noise events can closely resemble weak neutrino signals. However, unlike thermal noise, genuine neutrino signals, even at low SNR, are expected to be impulsive Askaryan signals. To distinguish between these dominant non-impulsive thermal noise backgrounds and potential neutrino events, we use several analysis variables, including signal impulsivity metrics, which are described in the following sections.

\subsection{Analysis Variables and Multivariate Selection}\label{sec:analysis_variables}
\par
\begin{figure}[htbp!]
  \centering
  \includegraphics[width=0.540\textwidth]{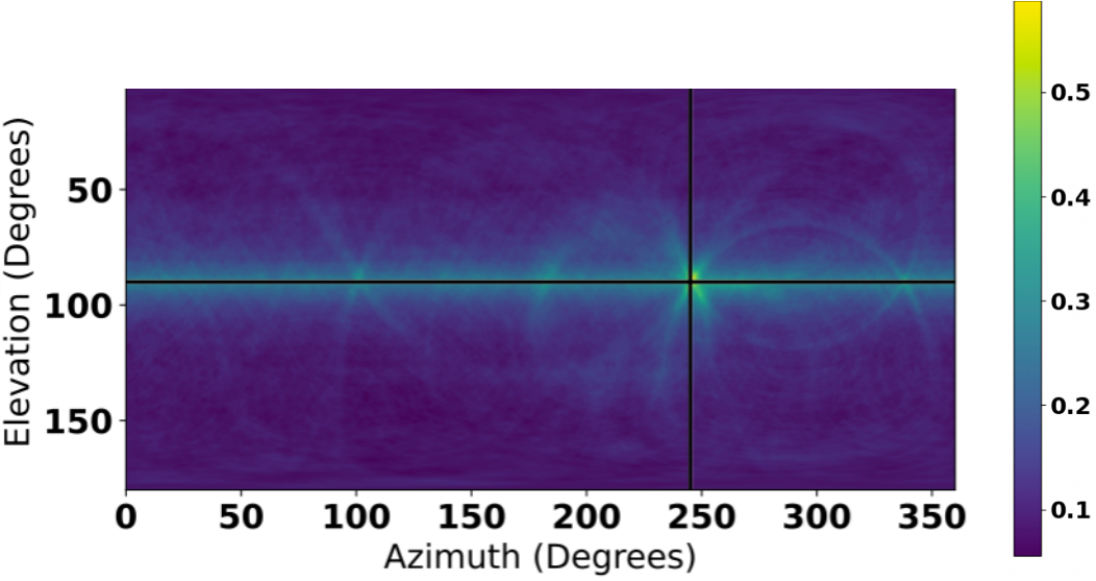}
  \hfill
  \includegraphics[width=0.400\textwidth]{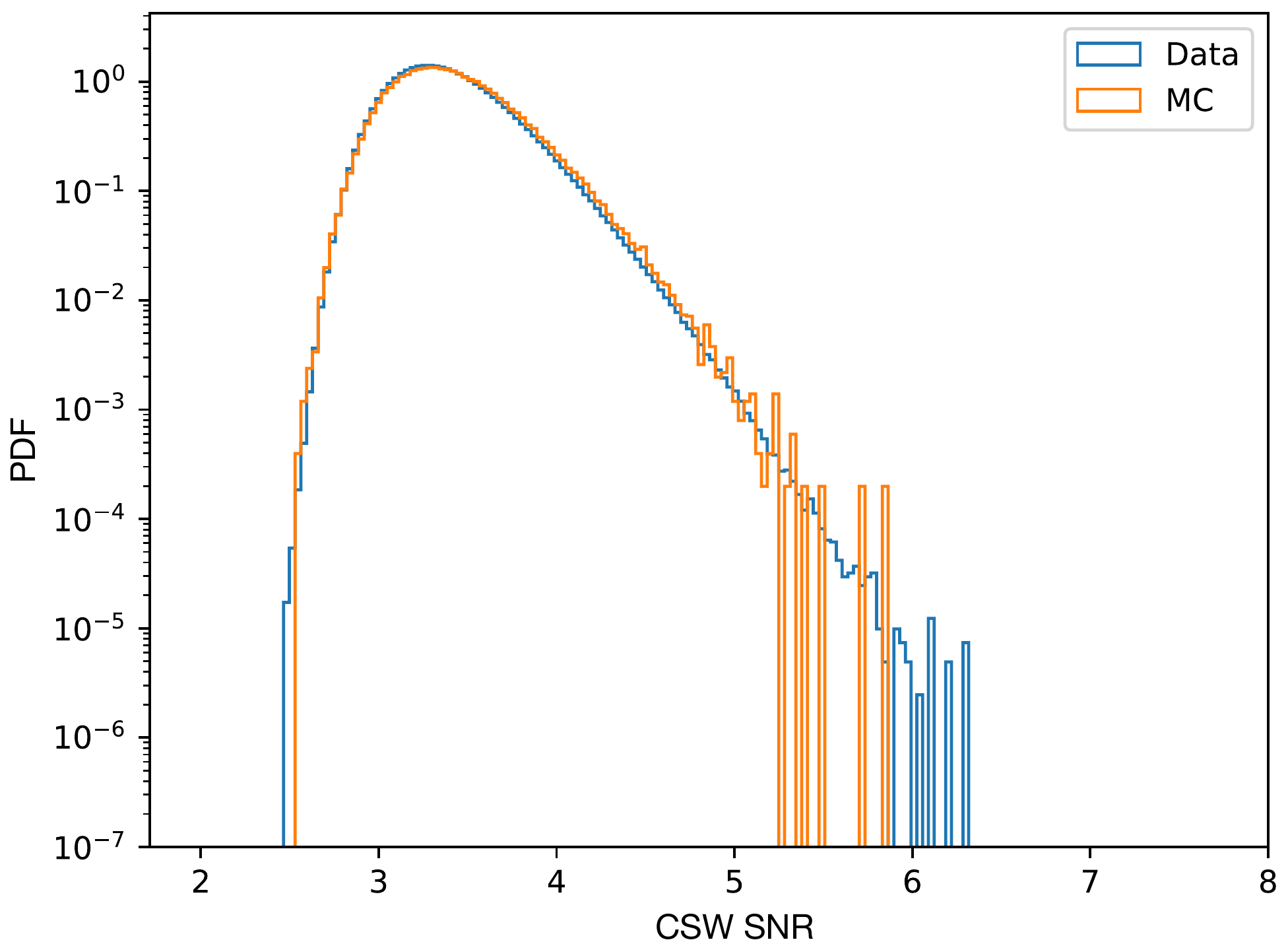}
  \caption{Left: Correlation map of a calibration pulser event showing reconstructed direction relative to the true source (black lines). Right: CSW SNR distribution for simulated thermal noise (orange, $\sim$100k) and real A5/PA hybrid data (blue, $\sim$10M), dominated by thermal noise.}
  \label{fig:lda}
\end{figure}

A central component of this analysis is reconstructing event arrival directions using two-dimensional correlation mapping in zenith ($\theta$) and azimuth ($\phi$) relative to the phased array string. For each event in the 10\% sample, we generate a $\theta$-$\phi$ correlation map by coherently summing signals across channels with relative time delays for plane-wave arrival directions. An example correlation map showing elevation versus azimuth for a local calibration pulser event is shown in Fig.~\ref{fig:lda} (left); elevation (el) and zenith ($\theta$) relate as $\theta = 90^\circ - \mathrm{el}$. The correlation peak indicates the most probable incoming signal direction.

\par

Using the time delays from the peak of the correlation map (Fig.~\ref{fig:lda}, left), we form a coherently summed waveform (CSW) for each event. From the correlation map and CSW, we extract key analysis variables, including maximum correlation value, radius of maximum correlation, reconstructed $\theta$ and $\phi$ angles, average SNR, CSW SNR, Hilbert envelope SNR, signal impulsivity, and others. These variables provide strong discrimination power between potential neutrino signals and backgrounds described in Section~\ref{sec:backgrounds}.

\subsection{Linear Discriminant Analysis and Sensitivity Projection}\label{sec:fisher_discriminant}
\par
To distinguish neutrino signals from the dominant thermal noise background, we apply linear discriminant analysis (LDA)~\cite{Fisher:1938}. This method combines multiple variables from Section~\ref{sec:analysis_variables} into a single discriminant (the LDA value), optimized to separate simulated neutrino events from background. Figure~\ref{fig:lda} (right) compares simulated thermal noise with A5/PA hybrid data, demonstrating excellent data-MC agreement and validating the underlying simulation framework. Neutrino events, simulated using the same configuration, are used to train the LDA and evaluate signal efficiency.
\begin{figure}[htbp!]
  \centering
  \includegraphics[width=0.510\textwidth]{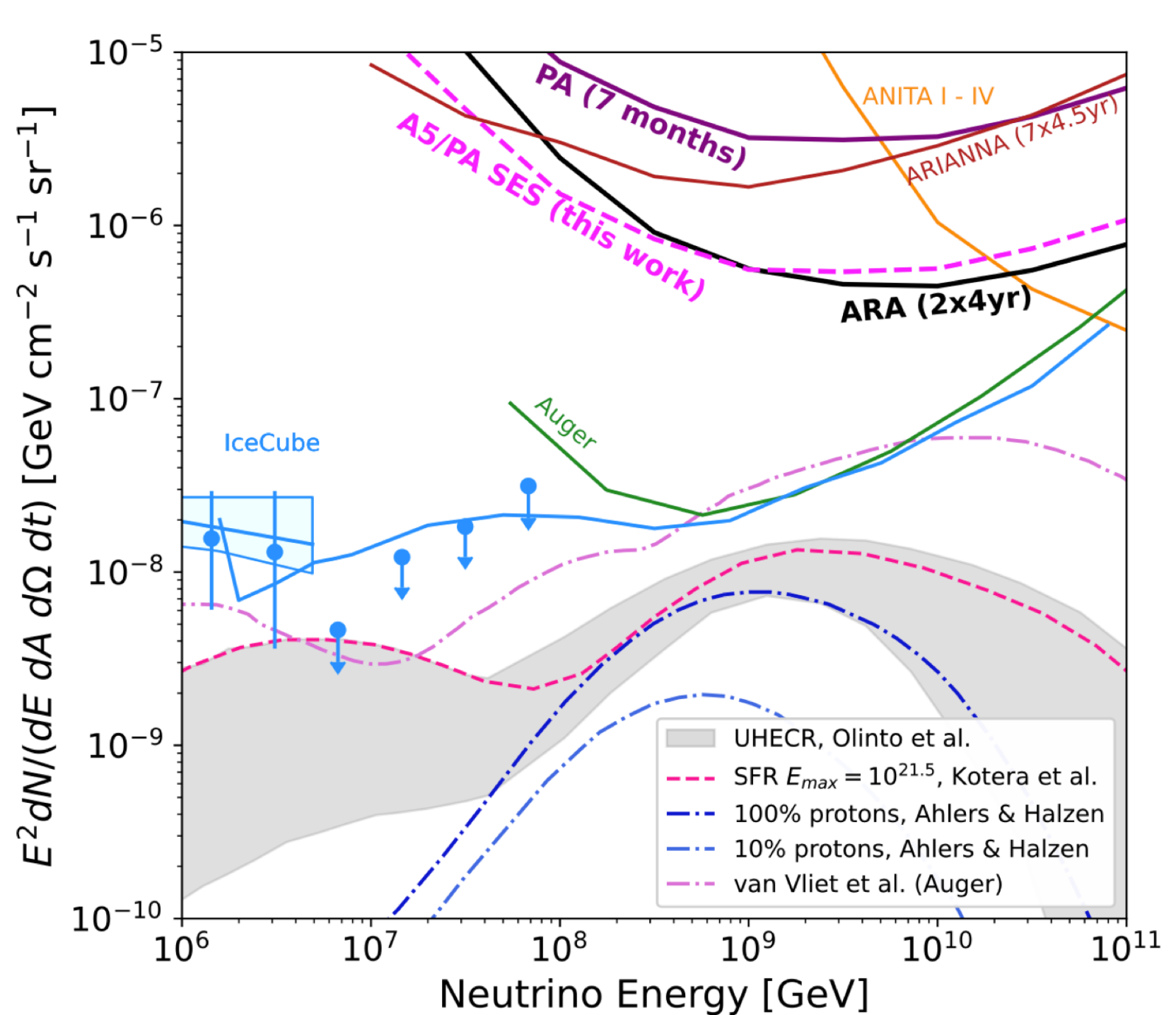}
  \caption{Projected single-event sensitivity of the A5/PA hybrid analysis (dashed), compared to previous ARA analyses and other experiments.}
  \label{fig:sensitivity}
  
\end{figure}

We optimize the LDA selection cut to yield the strongest expected 90\% Feldman-Cousins upper limit~\cite{PhysRevD.57.3873}, balancing signal efficiency and background leakage. Optimization assumes a neutrino flux consistent with the union of existing limits from IceCube~\cite{IceCube:2025ezc} and ANITA~\cite{ANITA:2019wyx}. The resulting cut is applied to the remaining 90\% of the data to derive the final flux limit.

\par
The projected single-event sensitivity (Fig.~\ref{fig:sensitivity}) surpasses previous ARA results, enabled by the hybrid system’s combination of low-threshold triggering and enhanced directional reconstruction. This approach improves background rejection and overall analysis efficiency. Final results will follow after unblinding the full dataset.

\section{Conclusion}\label{sec3}

\par
In recent years, the ARA Collaboration has demonstrated the advantages of a low-threshold beamforming trigger system for detecting ultrahigh energy neutrinos via radio techniques. This work presents a pioneering analysis that unifies advanced tools and combines data from a hybrid detector system integrating traditional ARA antennas with a phased array beamforming instrument. This first-of-its-kind analysis significantly improves ARA’s sensitivity in the sub-EeV energy range. The phased array’s efficient triggering enhances both trigger and analysis efficiencies~\cite{ARA:2022rwq}, while its combination with traditional antennas in a hybrid setup provides further gains.

\par
This analysis reflects years of dedicated effort by the ARA collaboration, incorporating improved background rejection techniques and enhanced detector and neutrino simulations, which lead to better data–Monte Carlo agreement and increased overall sensitivity. This hybrid approach will inform the design and analysis strategies of next-generation radio observatories, such as the proposed IceCube-Gen2 radio component. The recent 220~PeV neutrino candidate reported by KM3NeT~\cite{KM3NeT:2025npi} underscores the need for improved sensitivity below 1~EeV. The hybrid detector method presented here marks a crucial step forward, boosting the discovery potential of in-ice radio detectors in the near future.\\

\begingroup
\setstretch{0.25}   
\setlength{\bibsep}{0.75pt}
\bibliographystyle{ICRC}
\bibliography{references}

\providecommand{\href}[2]{#2}\begingroup\raggedright\begin{thebibliography}{10}

\bibitem{Greisen:1966jv}
K.~Greisen \href{http://dx.doi.org/10.1103/PhysRevLett.16.748}{{\em Phys. Rev. Lett.} {\bfseries 16} (1966) 748--750}.

\bibitem{Zatsepin:1966jv}
G.~T. Zatsepin and V.~A. Kuzmin {\em JETP Lett.} {\bfseries 4} (1966) 78--80.

\bibitem{Askaryan:1961pfb}
G.~A. Askar'yan {\em Zh. Eksp. Teor. Fiz.} {\bfseries 41} (1961) 616--618.

\bibitem{Prohira:2018mmv}
{\bfseries ANITA} Collaboration, S.~Prohira, A.~Novikov, P.~Dasgupta, {\em et~al.} \href{http://dx.doi.org/10.1103/PhysRevD.98.042004}{{\em Phys. Rev. D} {\bfseries 98} no.~4, (2018) 042004}.

\bibitem{IceCube-Gen2:2020qha}
{\bfseries IceCube-Gen2} Collaboration, M.~G. Aartsen {\em et~al.} \href{http://dx.doi.org/10.1088/1361-6471/abbd48}{{\em J. Phys. G} {\bfseries 48} no.~6, (2021) 060501}.

\bibitem{ARA:2019wcf}
{\bfseries ARA} Collaboration, P.~Allison {\em et~al.} \href{http://dx.doi.org/10.1103/PhysRevD.102.043021}{{\em Phys. Rev. D} {\bfseries 102} no.~4, (2020) 043021}.

\bibitem{ARA:2021wmr}
{\bfseries ARA} Collaboration, P.~Dasgupta and K.~Hughes \href{http://dx.doi.org/10.22323/1.395.1086}{{\em PoS} {\bfseries ICRC2021} (2021) 1086}.

\bibitem{Allison:2018ynt}
P.~Allison {\em et~al.} \href{http://dx.doi.org/10.1016/j.nima.2019.01.067}{{\em Nucl. Instrum. Meth. A} {\bfseries 930} (2019) 112--125}.

\bibitem{ARA:2022rwq}
{\bfseries ARA} Collaboration, P.~Allison {\em et~al.} \href{http://dx.doi.org/10.1103/PhysRevD.105.122006}{{\em Phys. Rev. D} {\bfseries 105} no.~12, (2022) 122006}.

\bibitem{Klein:2005di}
J.~R. Klein and A.~Roodman \href{http://dx.doi.org/10.1146/annurev.nucl.55.090704.151521}{{\em Ann. Rev. Nucl. Part. Sci.} {\bfseries 55} (2005) 141--163}.

\bibitem{arasim}
{The ARA Collaboration}. \url{https://github.com/ara-software/AraSim}.

\bibitem{ARA:2014fyf}
{\bfseries ARA} Collaboration, P.~Allison {\em et~al.} \href{http://dx.doi.org/10.1016/j.astropartphys.2015.04.006}{{\em Astropart. Phys.} {\bfseries 70} (2015) 62--80}.

\bibitem{Kotera:2010yn}
K.~Kotera, D.~Allard, and A.~V. Olinto \href{http://dx.doi.org/10.1088/1475-7516/2010/10/013}{{\em JCAP} {\bfseries 10} (2010) 013}.

\bibitem{DEVRIES201696}
K.~D. de~Vries, S.~Buitink, {\em et~al.} \href{http://dx.doi.org/10.1016/j.astropartphys.2015.10.003}{{\em Astropart. Phys.} {\bfseries 74} (2016) 96--104}.

\bibitem{ANITA:2018vwl}
{\bfseries ANITA} Collaboration, P.~W. Gorham {\em et~al.} \href{http://dx.doi.org/10.1103/PhysRevD.98.022001}{{\em Phys. Rev. D} {\bfseries 98} no.~2, (2018) 022001}.

\bibitem{Fisher:1938}
R.~A. Fisher {\em Annals of Eugenics} {\bfseries 8} no.~4, (1938) 376--386.

\bibitem{PhysRevD.57.3873}
G.~J. Feldman and R.~D. Cousins \href{http://dx.doi.org/10.1103/PhysRevD.57.3873}{{\em Phys. Rev. D} {\bfseries 57} (Apr, 1998) 3873--3889}.

\bibitem{IceCube:2025ezc}
{\bfseries IceCube} Collaboration, R.~Abbasi {\em et~al.}

\bibitem{ANITA:2019wyx}
{\bfseries ANITA} Collaboration, P.~W. Gorham {\em et~al.} \href{http://dx.doi.org/10.1103/PhysRevD.99.122001}{{\em Phys. Rev. D} {\bfseries 99} no.~12, (2019) 122001}.

\bibitem{KM3NeT:2025npi}
{\bfseries KM3NeT} Collaboration, S.~Aiello {\em et~al.} \href{http://dx.doi.org/10.1038/s41586-024-08543-1}{{\em Nature} {\bfseries 638} no.~8050, (2025) 376--382}. [Erratum: Nature 640, E3 (2025)].

\end{thebibliography}\endgroup
\endgroup

%

\clearpage

\section*{Full Author List: ARA Collaboration (June 30, 2025)}

\noindent
N.~Alden\textsuperscript{1}, 
S.~Ali\textsuperscript{2}, 
P.~Allison\textsuperscript{3}, 
S.~Archambault\textsuperscript{4}, 
J.J.~Beatty\textsuperscript{3}, 
D.Z.~Besson\textsuperscript{2}, 
A.~Bishop\textsuperscript{5}, 
P.~Chen\textsuperscript{6}, 
Y.C.~Chen\textsuperscript{6}, 
Y.-C.~Chen\textsuperscript{6}, 
S.~Chiche\textsuperscript{7}, 
B.A.~Clark\textsuperscript{8}, 
A.~Connolly\textsuperscript{3}, 
K.~Couberly\textsuperscript{2}, 
L.~Cremonesi\textsuperscript{9}, 
A.~Cummings\textsuperscript{10,11,12}, 
P.~Dasgupta\textsuperscript{3}, 
R.~Debolt\textsuperscript{3}, 
S.~de~Kockere\textsuperscript{13}, 
K.D.~de~Vries\textsuperscript{13}, 
C.~Deaconu\textsuperscript{1}, 
M.A.~DuVernois\textsuperscript{5}, 
J.~Flaherty\textsuperscript{3}, 
E.~Friedman\textsuperscript{8}, 
R.~Gaior\textsuperscript{4}, 
P.~Giri\textsuperscript{14}, 
J.~Hanson\textsuperscript{15}, 
N.~Harty\textsuperscript{16}, 
K.D.~Hoffman\textsuperscript{8}, 
M.-H.~Huang\textsuperscript{6,17}, 
K.~Hughes\textsuperscript{3}, 
A.~Ishihara\textsuperscript{4}, 
A.~Karle\textsuperscript{5}, 
J.L.~Kelley\textsuperscript{5}, 
K.-C.~Kim\textsuperscript{8}, 
M.-C.~Kim\textsuperscript{4}, 
I.~Kravchenko\textsuperscript{14}, 
R.~Krebs\textsuperscript{10,11}, 
C.Y.~Kuo\textsuperscript{6}, 
K.~Kurusu\textsuperscript{4}, 
U.A.~Latif\textsuperscript{13}, 
C.H.~Liu\textsuperscript{14}, 
T.C.~Liu\textsuperscript{6,18}, 
W.~Luszczak\textsuperscript{3}, 
A.~Machtay\textsuperscript{3}, 
K.~Mase\textsuperscript{4}, 
M.S.~Muzio\textsuperscript{5,10,11,12}, 
J.~Nam\textsuperscript{6}, 
R.J.~Nichol\textsuperscript{9}, 
A.~Novikov\textsuperscript{16}, 
A.~Nozdrina\textsuperscript{3}, 
E.~Oberla\textsuperscript{1}, 
C.W.~Pai\textsuperscript{6}, 
Y.~Pan\textsuperscript{16}, 
C.~Pfendner\textsuperscript{19}, 
N.~Punsuebsay\textsuperscript{16}, 
J.~Roth\textsuperscript{16}, 
A.~Salcedo-Gomez\textsuperscript{3}, 
D.~Seckel\textsuperscript{16}, 
M.F.H.~Seikh\textsuperscript{2}, 
Y.-S.~Shiao\textsuperscript{6,20}, 
S.C.~Su\textsuperscript{6}, 
S.~Toscano\textsuperscript{7}, 
J.~Torres\textsuperscript{3}, 
J.~Touart\textsuperscript{8}, 
N.~van~Eijndhoven\textsuperscript{13}, 
A.~Vieregg\textsuperscript{1}, 
M.~Vilarino~Fostier\textsuperscript{7}, 
M.-Z.~Wang\textsuperscript{6}, 
S.-H.~Wang\textsuperscript{6}, 
P.~Windischhofer\textsuperscript{1}, 
S.A.~Wissel\textsuperscript{10,11,12}, 
C.~Xie\textsuperscript{9}, 
S.~Yoshida\textsuperscript{4}, 
R.~Young\textsuperscript{2}
\\
\\
\textsuperscript{1} Dept. of Physics, Enrico Fermi Institute, Kavli Institute for Cosmological Physics, University of Chicago, Chicago, IL 60637\\
\textsuperscript{2} Dept. of Physics and Astronomy, University of Kansas, Lawrence, KS 66045\\
\textsuperscript{3} Dept. of Physics, Center for Cosmology and AstroParticle Physics, The Ohio State University, Columbus, OH 43210\\
\textsuperscript{4} Dept. of Physics, Chiba University, Chiba, Japan\\
\textsuperscript{5} Dept. of Physics, University of Wisconsin-Madison, Madison,  WI 53706\\
\textsuperscript{6} Dept. of Physics, Grad. Inst. of Astrophys., Leung Center for Cosmology and Particle Astrophysics, National Taiwan University, Taipei, Taiwan\\
\textsuperscript{7} Universite Libre de Bruxelles, Science Faculty CP230, B-1050 Brussels, Belgium\\
\textsuperscript{8} Dept. of Physics, University of Maryland, College Park, MD 20742\\
\textsuperscript{9} Dept. of Physics and Astronomy, University College London, London, United Kingdom\\
\textsuperscript{10} Center for Multi-Messenger Astrophysics, Institute for Gravitation and the Cosmos, Pennsylvania State University, University Park, PA 16802\\
\textsuperscript{11} Dept. of Physics, Pennsylvania State University, University Park, PA 16802\\
\textsuperscript{12} Dept. of Astronomy and Astrophysics, Pennsylvania State University, University Park, PA 16802\\
\textsuperscript{13} Vrije Universiteit Brussel, Brussels, Belgium\\
\textsuperscript{14} Dept. of Physics and Astronomy, University of Nebraska, Lincoln, Nebraska 68588\\
\textsuperscript{15} Dept. Physics and Astronomy, Whittier College, Whittier, CA 90602\\
\textsuperscript{16} Dept. of Physics, University of Delaware, Newark, DE 19716\\
\textsuperscript{17} Dept. of Energy Engineering, National United University, Miaoli, Taiwan\\
\textsuperscript{18} Dept. of Applied Physics, National Pingtung University, Pingtung City, Pingtung County 900393, Taiwan\\
\textsuperscript{19} Dept. of Physics and Astronomy, Denison University, Granville, Ohio 43023\\
\textsuperscript{20} National Nano Device Laboratories, Hsinchu 300, Taiwan\\

\section*{Acknowledgements}

\noindent
The ARA Collaboration is grateful to support from the National Science Foundation through Award 2013134.
The ARA Collaboration
designed, constructed, and now operates the ARA detectors. We would like to thank IceCube, and specifically the winterovers for the support in operating the
detector. Data processing and calibration, Monte Carlo
simulations of the detector and of theoretical models
and data analyses were performed by a large number
of collaboration members, who also discussed and approved the scientific results presented here. We are
thankful to Antarctic Support Contractor staff, a Leidos unit 
for field support and enabling our work on the harshest continent. We thank the National Science Foundation (NSF) Office of Polar Programs and
Physics Division for funding support. We further thank
the Taiwan National Science Councils Vanguard Program NSC 92-2628-M-002-09 and the Belgian F.R.S.-
FNRS Grant 4.4508.01 and FWO. 
K. Hughes thanks the NSF for
support through the Graduate Research Fellowship Program Award DGE-1746045. A. Connolly thanks the NSF for
Award 1806923 and 2209588, and also acknowledges the Ohio Supercomputer Center. S. A. Wissel thanks the NSF for support through CAREER Award 2033500.
A. Vieregg thanks the Sloan Foundation and the Research Corporation for Science Advancement, the Research Computing Center and the Kavli Institute for Cosmological Physics at the University of Chicago for the resources they provided. R. Nichol thanks the Leverhulme
Trust for their support. K.D. de Vries is supported by
European Research Council under the European Unions
Horizon research and innovation program (grant agreement 763 No 805486). D. Besson, I. Kravchenko, and D. Seckel thank the NSF for support through the IceCube EPSCoR Initiative (Award ID 2019597). M.S. Muzio thanks the NSF for support through the MPS-Ascend Postdoctoral Fellowship under Award 2138121. A. Bishop thanks the Belgian American Education Foundation for their Graduate Fellowship support.

\end{document}